			   \documentclass[
       ,final            
  ]
  {aipproc}

\layoutstyle{8x11double}
\usepackage{amsmath}
\newcommand{\tmc}{T_\mathrm{mc}}
\newcommand{\texp}{\tau_\mathrm{exp}}

\begin{document}

\title{Finite size effects in the specific heat of glass-formers}


\author{L.~A.~Fernandez}{
address={Departamento de F\'isica Te\'orica I, Universidad Complutense,
Avenida Complutense, 28040 Madrid, Spain}
}

\author{V.~Mart\'in-Mayor}{
address={Departamento de F\'isica Te\'orica I, Universidad Complutense,
Avenida Complutense, 28040 Madrid, Spain}
}

\author{P.~Verrocchio}{
address={Dipartimento di Fisica, Universit\`a di Trento, via Sommarive 14,
38050 Povo, Trento, Italy} 
}

\begin{abstract}
  We report clear finite size effects in the specific heat and in the
  relaxation times of a model glass former at temperatures
  considerably smaller than the Mode Coupling transition. A crucial
  ingredient to reach this result is a new Monte Carlo algorithm which
  allows us to reduce the relaxation time by two order of magnitudes.
  These effects signal the existence of a large correlation length in
  static quantities.
\end{abstract}

\maketitle


\section{Introduction}
When compared to experiments, numeric simulations of glass-forming
systems present severe limitations. In fact, while the actual physical
origin of the phenomenon termed generically as {\em glass transition}
is still a matter of debate, its investigation is limited by the
practical impossibility to reach thermodynamic equilibrium when the
relaxation time $\tau$ of a sample glass in a laboratory becomes of
order of $10^2$ seconds or when the $\tau$ of a model glass in a
modern computer becomes of $\sim 10^{-8}$ seconds.  Thus the glass
temperature $T_g$ is conventionally defined in experiments and numeric
simulations cannot get very close to it~\footnote{e.g. in $SiO_2$
Molecular Dynamics simulation have been performed only down to $T \sim
2200 K$ while $T_g=1500 K$.}.  Furthermore, the statistical accuracy
in experiments is fairly larger, because the time-length ${\cal L}$ of
typical experiments is many decades larger than in simulations.
Indeed, statistical errors decrease roughly as $1/\sqrt{{\cal
N}_{ind}}$ where the number of {\em independent} configurations ${\cal
N}_{ind}$ is given by ${\cal N}_{ind} \sim {\cal L}/\tau$. This
problem is particularly serious in the study of time correlators,
whose peculiar behavior is one of the most clear signature of the
glass transition.  For every observable, $O$, the normalized time
correlator is
\begin{equation}
C_O(t)=\frac{\langle O (0) O(t)\rangle-\langle O\rangle^2}{\langle
O^2\rangle-\langle O\rangle^2}\,.\label{CODEF}
\end{equation}
The region $C \sim 0$ is very interesting, since it allows to estimate
the relaxation time (see below). On the other hand, in that region the
relative error from a {\em single} measurement grows as $C^{-1}$ hence
a very large ${\cal N}_{ind}$ is needed to get some significant signal
over the noise. As a consequence, most of previous numeric work was
confined to $C>0.1$, while in experiments one is able to explore far
lesser values of $C$.

Despite those limitations, numeric simulations turn out to be quite
useful in many respects, mostly in the key issue of identifying the
quantity suffering the largest spatial fluctuations at the glass
transition.\cite{tarjus96,donati02,cugliandolo:2003,biroli:2004,whitelam:2004}
In fact, unlike standard phase transitions the equilibration time of
glass-formers (supercooled liquids, polymers, proteins,
superconductors, etc.) diverges without dramatic changes in their
structural
properties\cite{DeBenedetti97,angell:2000,DeBenedettiStillinger01},
challenging the interpretation of the slowing down as a critical
phenomenon\cite{ZINNJUSTIN}. A possible way out has been found in the
{\em heterogeneous dynamic scenario}, which postulate that {\em local}
time correlators (i.e where $O$ is a local quantity) are correlated
over diverging length scales. The possible advantages of numeric
simulations with respect of experiment are twofold. First, they are
more suitable to investigate local quantities. Second, they allow an
easier implementation of Finite Size Scaling techniques, a powerful
tool to detect diverging correlation
length.\cite{cardy,FSSBOOK,BERTHIER03} New, fast algorithms for
numerical simulation are therefore needed in order to to get as close
as possible to the experimental regime, while retaining all the
advantages provided by numerics.

\section{Translational invariant quantities}

In this numeric work, we wish to ask the following question: Are there
static observables whose correlation length $\xi$ diverge at the glass
transition?  (Note that experimental clues are not available).  In
other words we wonder if a standard second order phase transition is
behind the glass transition. We borrow from the dynamic heterogeneity
scenario the hypothesis that a diverging correlation length would show
up in four-particle correlation function and focus on translational
invariant quantities.

Little attention has been paid so far to such quantities. In fact
experiments (e.g. elastic scattering for the spatial correlations and
dielectric response for the dynamic behavior) and simulations
typically investigate quantities which are not invariant under an
uniform displacement of the coordinates, $\vec r_i =\vec r_i +\vec
\Delta$, namely the density fluctuations of generic momentum $\vec Q$:
\begin{equation}
\rho(\vec Q) = \frac{\sigma_0^{-3}}{N}\sum_{j=1}^N \mathrm{e}^{\mathrm{i} \vec
r_j\cdot \vec Q},
\label{DENSITY}
\end{equation}
where $\rho(0)$ is the particle density and it is constant because of
mass conservation. Density fluctuations at small $\vec Q$ are surely
slow modes of the system, their time-correlator decaying exponentially
with characteristic time $\tau \sim D^{-1} Q^{-2}$ ($D$ is the
diffusivity). We have no guarantee whatsoever though that they are the
only relevant slow modes.

How to expand the space of observables investigated taking into
account even quantities which preserve the translational symmetry?
For the sake of definiteness we limit to the $NVT$ ensemble and
introduce energy density fluctuations:
\begin{equation}
\rho_e(\vec Q)= \frac{\sigma_0^{-3}}{2N}
\sum_{j,k\neq j}
\mathrm{e}^{\mathrm{i} \vec r_j\cdot \vec Q} V_{kj}.
\end{equation}
where $V_{kj}$ is the interaction energy $V({\vec r_k -\vec r_j})$
between particles $k$ and $j$. Note that $\rho_e(0)$ is the
non-conserved potential energy density, which we call $e$ (the
internal energy is then $\frac32 k_\mathrm{B} T +\langle e\rangle$).
A general way to obtain translationally invariant observables consists
in multiplying densities with wave-vectors $\vec q$ and $-\vec q$:
\begin{equation}
{\cal S}(\vec q) = \rho(\vec q) \rho(-\vec q)\ ,\quad {\cal S}_e(\vec q)=
\rho_e(\vec q) \rho_e(-\vec q)\,.\label{SQDEF}
\end{equation}
We see then that the correlation functions of these translational
invariant quantities involve the positions of four particles. Moreover
the mean value of ${\cal S}(\vec q)$ is the static structure factor
$S(\vec q)$, while $\langle {\cal S}_e(0)\rangle$ is related to the
constant-volume specific heat, $C_V$ as $T^2C_V= N
(\sigma_0)^6\left[\langle {\cal S}_e(0)\rangle - \langle e\rangle
^2\right]\,.$

\section{The local swap Monte Carlo algorithm}

\begin{table}
\begin{tabular}{lrrrrrr}
\hline
    \tablehead{1}{l}{b}{$T$} 
  & \tablehead{1}{r}{b}{$2.13\,\tmc$}
  & \tablehead{1}{r}{b}{$1.08\,\tmc$}
  & \tablehead{1}{r}{b}{$1.02\,\tmc$}
  & \tablehead{1}{r}{b}{$0.97\,\tmc$}
  & \tablehead{1}{r}{b}{$0.92\,\tmc$} 
  & \tablehead{1}{r}{b}{$0.89\,\tmc$} \\
\hline
$\beta$ & 0.99(3) & 0.99(8) & 0.98(12) & 0.89(13) & 0.83(4) & 0.93(12) \\ 
\hline
\end{tabular}
\caption{Stretching exponent $\beta$ of the time correlator of the energy at 
different temperatures. The fit to a stretched exponential form 
has been performed in the region $C < 0.1$}
\label{tab:a}
\end{table}

We study a 50\% mixture of particles interacting through the pair
potential
$V_{\alpha\beta}(r)=\epsilon[(\sigma_\alpha+\sigma_\beta)/r]^{12} +
C_{\alpha\beta}$, where $\alpha,\beta=A,B$, with a cutoff at
$r_\mathrm{c}=\sqrt{3}\sigma_0$ ($\sigma_0$ is the unit length). The
choice $\sigma_B=1.2\sigma_A$ hampers crystallization. We impose
$(2\sigma_A)^3+2(\sigma_A+\sigma_B)^3+(2\sigma_B)^3=4\sigma_0^3$.
Constants $C_{\alpha \beta}$ are chosen to ensure continuity at
$r_\mathrm{c}$. The simulations are at constant volume, with particle
density fixed to $\sigma_0^{-3}\,$ and temperatures in the range
$[0.897 T_\mathrm{mc}, 10.792 T_\mathrm{mc}]$, where $T_\mathrm{mc}$
is the Mode Coupling temperature\cite{goetze:1992}.  We use periodic
boundary conditions on a box of size $L$ (which discretizes momenta in
units of $q_\mathrm{min}=2\pi/L\,$) in systems with $N=512,1024,2048$
and $4096$ particles.  For argon parameters, $\sigma_0=3.4$\AA,
$\epsilon/k_B=120$K and $T_\mathrm{mc}=26.4$K.

We modify the Grigera-Parisi swap algorithm\cite{Grigera01} to make it
{\em local}, in order to keep the algorithm in the dynamic
Universality Class\cite{ZINNJUSTIN} of standard Monte Carlo (MC).  The
elementary MC step is either (with probability $p$) a single-particle
displacement attempt or (with probability $1-p$) an attempt to {\em
swap} particles. The swap is made by picking a particle at random and
trying to interchange its position with that of a particle of opposite
type, chosen at random {\em among those at distance smaller than $0.6
r_\mathrm{c}$}.  Detailed balance requires that the Metropolis test
include not only the energy variation but the change in the number of
neighbors. The swap acceptance is independent of system size and
grows from 0.74\% at 0.9$T_\mathrm{mc}$ up to 6\% at
$2T_\mathrm{mc}$. In this work we use $p=0.5$ (named {\em local swap}
from here on) and $p=1.0$ (named standard MC). The time unit $t_0$ is
$N/p$ elementary steps.

\begin{figure}
  \includegraphics[height=.3\textheight,angle=270]{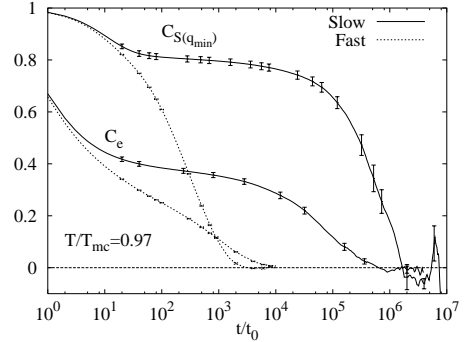}
  \caption{Correlators for two translationally invariant quantities: the
    potential energy density, $e$, and the square of the density
    fluctuation, ${\cal S}(q_\mathrm{min})$, at the minimal momentum
    allowed by the boundary conditions ($1024$ particles, below
    $T_\mathrm{mc}$). The local swap algorithm decorrelates faster
    than standard MC.\protect{\label{FIG1}}}
\end{figure}

In fig.~(\ref{FIG1}) we show that the time-correlator of the local
swap is in general different from the one of the standard MC. The main
difference is given by the absence of the cage effect when the
particles are allowed to swap.  This is seen by the absence of the
plateau in the time correlator in the swap dynamics. This implies that
Mode Coupling transition is rather ill-defined within the swap
dynamics. In fact, at mean field level Mode Coupling singularity is
seen as due to the divergence of the caging time
scale~\cite{goetze:1992}.
The Theory of Critical Phenomena suggests that the decay of time
correlators of two different dynamics sharing both the local nature
and the conservation laws is given by the same fuction of the
correlation length $\xi$ (up to non universal numerical
prefactors). They are said in fact to belong to the same dynamic
Universality Class~\cite{ZINNJUSTIN}. Since at {\em very long} times
that decay is expected to be exponential it is possible to define the
{\em exponential autocorrelation time}~\cite{SOKALLECTURES} which
identifies the longest characteristic relaxation time $\tau$:
\begin{equation}
C_O(t)\underset{t\to\infty}\longrightarrow
\mathrm{e}^{-t/\tau}\,.\label{TAUEXP}
\end{equation}
where
\begin{equation}
\tau \sim \xi^z
\label{ZETA}
\end{equation}
the dynamic exponent $z$ depending only the on the Class of the
dynamics.

\begin{figure}
  \includegraphics[height=.3\textheight,angle=270]{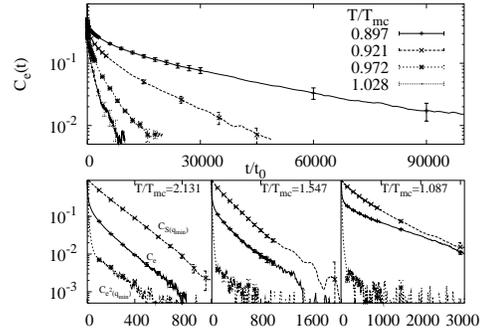}
  \caption{Time correlators for $N=1024$ particles ($N=2048$ at 
    $T/T_\mathrm{mc}=0.897$) with the swap algorithm.  Top panel:
    temperature variation of the correlator of $e$, close to
    $T_\mathrm{mc}$.  Bottom panels: correlators of several
    translational invariant quantities.  Asymptotically, three
    parallel straight lines should be found, with slope $-1/\texp$. At
    the highest temperature, (bottom,left), this is clearly observed.
    At lower temperatures (bottom,right) we do not have enough
    independent measurements to see clearly this common slope.
    \protect{\label{FIG2}}}
\end{figure}

In a critical phenomenon one expects that eventually all the
observables with the same symmetry decay with the same rate $1/\tau$
since they are all coupled to the slowest mode. The order parameter is
supposed to be such slowest mode and its decay is purely exponential.
In fig.~(\ref{FIG2}) we show the large time decay of different
translational invariant quantities obtained with the local swap. At
high temperatures the exponential time is clearly the same, on the
other hand at lower temperatures this is more difficult to show, since
the point where in log-linear representation all the curves become
parallel moves at lower and lower values of $C$.
Actually, in literature the time decay of glass-former is more often
described by stretched exponentials like $C_O(t) \sim
\exp{-\left(t/\tau\right)}^\beta$, with $\beta <1$. The stretching
would arise from the contribution of many regions relaxing with
different time scales. For the quantities we study in this work the
stretched exponential fit nicely the time correlators only at
intermediate times while, as we show in table~\ref{tab:a}, in the
asymptotic region $C<0.1$ the decay is in fact purely exponential.
It would be very difficult to get this result with standard algorithm
since at large $t$ the correlator is {\em very} noisy.
\begin{figure}
\includegraphics[angle=270,width=\columnwidth]{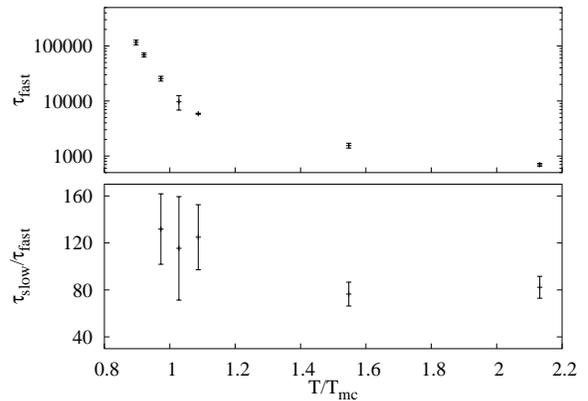}
\caption{Top: exponential time $\tau_{fast}$ of the translational
  invariant quantities with the swap dynamics for $N=1024$ particles
  ($N=2048$ at $T/T_\mathrm{mc} = 0.897$). Bottom: the ratio of the
  exponential times for standard MC ($\tau_{slow}$) and local swap
  dynamics ($\tau_{fast}$).  \protect{\label{FIG3}}}
\end{figure}

The advantage of local swap over standard MC is not limited to the
lack of the cage effect. This would be of no use if the exponential
time of the two dynamics was the same. We see instead in
Fig.~(\ref{FIG3}) that, although the $\tau$ of the local swap seems to
diverge, it remains always two order of magnitude smaller than the
exponential time for the standard MC. Because of eq.~(\ref{ZETA}) the
fact that this ratio is roughly constant at all temperatures suggests
that $z$ is the same for both dynamics and then that they belong to
the same Universality class. As anticipated in the introduction,
reducing $\tau$ allows us to obtain results with an higher statistical
accuracy. We show in fact in table~\ref{tab:b} some parameters
describing a selection of recent numeric simulations using standard
algorithms (molecular dynamics) to be compared in table~\ref{tab:c}
with the results that we were able to obtain with the local swap
algorithm.

\begin{table}
\begin{tabular}{lrrr}
\hline
    \tablehead{1}{l}{b}{Paper}
  & \tablehead{1}{r}{b}{Size\tablenote{Number of particles.}}
  & \tablehead{1}{r}{b}{Lowest T\tablenote{Lowest temperature where the
  authors equilibrated the system.}}
  & \tablehead{1}{r}{b}{Time\tablenote{simulation length in units of
  relaxation time.}}   \\
\hline
Yu \& Carruzzo \cite{CARRUZZO04} & $512$ & $1.01 T_\mathrm{mc}$ & $10$
\tablenote{the main finding of this paper is that the specific
heat yields a thermalization time about 100 times larger than the
alpha correlation time. In our simulations, this is the relaxation
time considered.} \\ 
Berthier\cite{BERTHIER04} & $1372$ & $ 0.97 T_\mathrm{mc}$ & $100$ \\ 
Lacevic et al.\cite{LACEVIC03} & $8000$ & $1.03 T_\mathrm{mc}$ & $222$ \\
\hline
\end{tabular}
\caption{Statistical accuracy in the most recent Molecular Dynamics
studies of simple glass-formers}
\label{tab:b}
\end{table}

\begin{table}
\begin{tabular}{lrr}
\hline
  \tablehead{1}{l}{b}{Temperature}
  & \tablehead{1}{r}{b}{Size}
  & \tablehead{1}{r}{b}{Time} \\
\hline
$0.97  T_\mathrm{mc}$ & $1024$ & $100000$ \\
$0.897 T_\mathrm{mc}$ & $1024$ & $27000$  \\
$0.897 T_\mathrm{mc}$ & $2048$ & $3000$   \\
$0.897 T_\mathrm{mc}$ & $4096$ & $400$   \\
\hline
\end{tabular}
\caption{Statistical accuracy reached by means of the new local swap}
\label{tab:c}
\end{table}

\section{Finite size effects}

Since the possible order parameter of the glass transition is unknown,
a direct measure of the correlation length is difficult, but one may
detect it indirectly through Finite-Size Scaling\cite{FSSBOOK} by
performing numeric simulations of system with different sizes.  In
fact, in the critical region, quantities related with fluctuations
(such as the specific heat) depend on the size of the system. Under
the hypothesis that the order parameter belongs to the class of
translational invariant quantities, its exponential time $\tau$ has to
grow with system size as a power law (the exponent being $z$) until
the correlation length becomes smaller than the linear size of the
system.  Thus, in general, Eq.~(\ref{ZETA}) holds in the thermodynamic
limit while it has to be modified in systems of finite
size~\cite{SOKALLECTURES}:
\begin{equation}
\tau \sim \max(L,\xi)^z.
\label{FFS}
\end{equation}
Another quantity of interest is the specific heat $C_v$. As a matter
of fact it must show a critical behavior if a second order phase
transition is the responsible for the glass transition. In this case
the approach to the thermodynamic value is rather expected to be
exponential in the regime where $L<\xi$.

\begin{figure}
\includegraphics[angle=270,width=\columnwidth]{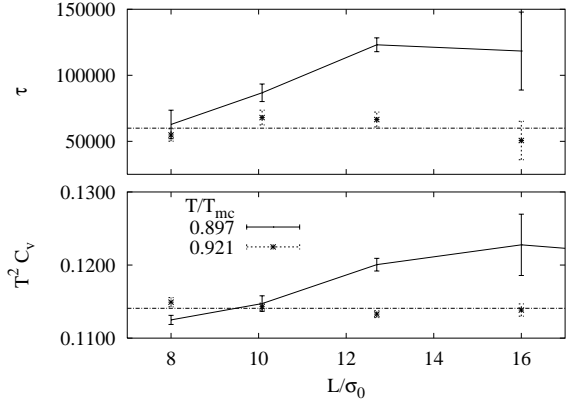}
\caption{Top: size dependence of the exponential time at the two 
  smallest temperatures studied, $T/T_\mathrm{mc}=0.921, 0.897$.
  Bottom: size dependence of the specific heat.
  \protect{\label{FIG4}}}
\end{figure}

In Fig.~(\ref{FIG4}) we show then the relaxation time and the specific
heat dependence on the size of the simulation box, $L$.  The main
result is that at the smallest studied temperature $T\!=\!0.897
T_\mathrm{mc}$, a noticeable growth of both quantities with $L$ is
found up to $L \sim 12.7\sigma_0$ ($\sim 4.3$nm for argon parameters)
which is our estimation of the correlation length $\xi$. The quantity
defined by eq.~(\ref{TAUEXP}) represents the {\em largest} relaxation
time, which is the one affected by criticality. Of course, different
time-scales (e.g. the plateau of intermediate scattering functions)
might display a different dependence on $L$~\cite{KIM00}.

The correlation length that we have found is
comparable with the experimental domain size for cooperative
dynamics\cite{EDIGER00,ISRAELOFF00}, and well above previous
simulations\cite{BERTHIER04}.  Aiming to estimate to dynamic exponent
$z$ one needs the value of $\xi$ at larger temperatures, however up to
$T\!=\!0.921 T_\mathrm{mc}$ no finite-size effects are detected for
both quantities. We believe that one should study smaller system to
find them, but unfortunately they crystallize quickly below $2.13
T_\mathrm{mc}$.
Taking $8 \sigma_0$ as upper bound for $\xi$ at $T\!=\!0.921
T_\mathrm{mc}$ and utilizing the values of $\tau$ at the two smallest
temperature we derive from Eq.~(\ref{ZETA}) $z \sim 1.2$ as a rough
estimate of the upper bound for the dynamic exponent.

\section{Conclusions}
Our findings are relevant in a twofold way.
\begin{description}
\item[Fast Monte Carlo algorithm] We have shown that fast algorithms
  may open for numeric simulation a window on the phenomenology of the
  glass transition that was previously available only to experiments.
  On the other hand, the local swap dynamics introduced works quite
  well with the simple model studied here but it becomes less
  effective when the parameters of the model change. In a similar soft
  sphere model~\cite{CARRUZZO04} with $\sigma_B=1.4\sigma_A$, the swap
  acceptance reduces of two order of magnitudes. 
\item[Finite size effects] Having looked for a diverging correlation
  length we must at the moment content with a static correlation
  length that, close to the glass transition, is an order of magnitude
  larger than the range of the interactions. This challenges the
  common view that in glasses only dynamic quantities become
  correlated at large scales.  More importantly, the study of the
  dynamics of translationally invariant quantities appears as a
  challenge to experimentalists.  While measurements of the frequency
  dependence of the specific heat\cite{CARRUZZO04,NAGEL85} are an
  appealing possibility to estimate the potential energy relaxation
  time, the correlation-length could be studied by Finite-Size Scaling
  of the specific-heat and of relaxation times in films or in larger
  pores than previously used to confine glass-formers. In fact
  experiments in films\cite{FSS-FILMS} and nanopores\cite{FSS-PORES}
  show that the glass transition changes in samples with one or more
  dimensions of nanometric scale~\footnote{although it is difficult to
  disentangle Finite-Size Scaling from the effects of the interaction
  with the substrate.}. Interestingly, the specific-heat of toluene
  confined on pores of diameter 8.7\,nm \cite{FSS-PORES}, close to its
  glass temperature, is significantly smaller than for bulk toluene,
  which could signal a correlation length well above the nanometric
  scale.
  
  At a qualitative level one may note that a second order phase
  transition is not the only scenario predicting Finite Size effects
  in the specific heat. In fact, in the random first order transition
  picture of the glass transition~\cite{MOSAIC:1989} recently revised
  by Biroli and Bouchaud~\cite{biroli:2004b} the contribution to the
  specific heat from the configurational entropy $s_c(T)$ (the
  logarithm of the number of metastable states) is sizeable only when
  the system is large enough to contain many regions in different
  states. This happens when
  \begin{equation}
    L^d \, s_c(t) > 1.
   \label{MOSAIC}
  \end{equation}
  To discriminate between the two interpretations a direct measurement
  of the correlation length of the energy would be needed.
  Nevertheless, in~\cite{fernandez:2005} we have shown some evidence
  for a singular behavior of the specific heat, consistent with the
  second order phase transition interpretation.

\end{description}

\begin{theacknowledgments}
  We thank G. Biroli for pointing out that the random first order
  transition picture may explain finite size effects in the specific
  heat. P.V. was supported by the EC (contract MCFI-2002-01262).  We
  were partly supported by MEC (Spain), through contracts
  BFM2003-08532, FIS2004-05073 and FPA2004-02602.  The total CPU time
  devoted to the simulation (carried out at BIFI PC clusters) amounts
  to 10 years of 3 GHz Pentium IV.
\end{theacknowledgments}



\bibliographystyle{aipproc}   


\bibliography{sendai}

\end{document}